\documentclass[figure]{pasj00}
\draft

\Received{$\langle$reception date$\rangle$}
\Accepted{$\langle$acception date$\rangle$}
\Published{$\langle$publication date$\rangle$}
\SetRunningHead{Ohashi et al.}{The chemical variations in Vela C}

\usepackage{lscape}

\begin{document}
\title{Chemical evolution of the HC$_3$N and N$_2$H$^+$ molecules in dense cores of the Vela C giant molecular cloud complex}
\author{Satoshi \textsc{Ohashi},\altaffilmark{1}
Ken'ichi \textsc{Tatematsu},\altaffilmark{2,3}
Kosuke \textsc{Fujii},\altaffilmark{1}
Patricio \textsc{Sanhueza},\altaffilmark{2}\\
Quang \textsc{Nguyen Luong},\altaffilmark{2}
Minho \textsc{Choi},\altaffilmark{4}
Tomoya \textsc{Hirota},\altaffilmark{2,3}
and
Norikazu \textsc{Mizuno}\altaffilmark{1,2}
}
\altaffiltext{1}{Department of Astronomy, The University of Tokyo, Bunkyo-ku, Tokyo 113-0033}
\altaffiltext{2}{National Astronomical Observatory of Japan, 
2-21-1 Osawa, Mitaka, Tokyo 181-8588}
\altaffiltext{3}{Department of Astronomical Science, SOKENDAI (The Graduate University for Advanced Studies), 
2-21-1 Osawa, Mitaka, Tokyo 181-8588}
\altaffiltext{4}{Korea Astronomy and Space Science Institute, 
Daedeokdaero 776, Yuseong, Daejeon 305-348, South
Korea}

\email{satoshi.ohashi@nao.ac.jp
}

\maketitle

\begin{abstract}
We have observed the HC$_3$N ($J=10-9$) and N$_2$H$^+$ ($J=1-0$) lines toward the Vela C molecular clouds with the Mopra 22 m telescope to study chemical characteristics of dense cores. The intensity distributions of these molecules are similar to each other at an angular resolution of  53$\arcsec$, corresponding to 0.19 pc suggesting that these molecules trace the same dense cores. We identified 25 local peaks in the velocity-integrated intensity maps of the HC$_3$N and/or N$_2$H$^+$ emission. Assuming LTE conditions, we calculated the column densities of these molecules and  found  a tendency that N$_2$H$^+$/HC$_3$N abundance ratio seems to be low in starless regions while it seems to be high in star-forming regions, similar to  the tendencies in the NH$_3$/CCS, NH$_3$/HC$_3$N, and N$_2$H$^+$/CCS abundance ratios found in previous studies of dark clouds and the Orion A GMC. We suggest that carbon chain molecules, including HC$_3$N, may trace chemically young molecular gas and N-bearing molecules, such as N$_2$H$^+$, may trace later stages of chemical evolution in the Vela C molecular clouds. 
It may be possible that the N$_2$H$^+$/HC$_3$N abundance ratio of $\sim$ 1.4 divides the star-forming and starless peaks in the Vela C, although it is not as clear as those in NH$_3$/CCS, NH$_3$/HC$_3$N, and N$_2$H$^+$/CCS for the Orion A GMC. 
This less clear separation may be caused by our lower spatial resolution or the misclassification of star-forming and starless peaks  due to the larger distance of the Vela C. It might be also possible that the HC$_3$N ($J=10-9$) transition is not a good chemical evolution tracer compared with CCS ($J=4-3$ and $7-6$) transitions. 

\end{abstract}

\section{Introduction}

In nearby cold dark clouds, carbon-chain molecules such as HC$_3$N and CCS, and N-bearing molecules such as NH$_3$ and N$_2$H$^+$, are thought to be good tracers of chemical evolution (e.g., \cite{hir92,suz92,ben98,mae99,hir02,hir09}). 
Carbon-chain molecules are produced in chemically young evolutionary stages by ion-molecular reactions and depleted through both the gas-phase reactions and the adsorption onto dust grains (\cite{aik01}, \cite{hir10}).
On the other hand, N$_2$H$^+$ is produced in later evolutionary stages and is hardly depleted compared with other species because the nitrogen molecule N$_2$ (the precursor of N$_2$H$^+$) is produced in late stages of the gas-phase chemistry (\cite{aik01}). 
Therefore, it is suggested that carbon-chain molecules, CCS and HC$_3$N, trace 
the early chemical 
evolutionary stage, whereas N-bearing molecules, 
NH$_3$ and N$_2$H$^+$, trace later stages of the chemical evolution.
\citet{suz92} observed CCS ($J_N=4_3-3_2, J_N=2_1-1_0$), HC$_3$N ($J=5-4$), and NH$_3$ ($J,K=1,1$) emission toward nearby cold dark cores and found that CCS and HC$_3$N are abundant in starless cores while NH$_3$ is abundant in star-forming cores. \citet{ben98} observed the CCS ($J_N=4_3-3_2$) and N$_2$H$^+$ ($J=1-0$) emission toward dense cores in nearby cold dark clouds, and  showed that  the column density of CCS is anticorrelated with that of N$_2$H$^+$ among the cores. Therefore, it is established that the column density ratios of N-bearing molecules to carbon chain molecules can be indicators of chemical evolution, in cold dark clouds.

On the other hand, the chemical evolution of molecular dense cores located in giant molecular clouds (GMCs) has not been extensively explored in comparison with that of nearby cold dark clouds. The dense cores in nearby dark clouds are known to be cold ($\sim10$ K), less turbulent, and show isolated low mass star-forming region, while those in GMCs are warmer, turbulent, and often show the formation of star clusters. 
Because most stars in the Galaxy are formed in GMCs, it is of great interest to know whether or not these molecules are indicators of chemical evolution in GMCs.
It has been shown that the column density ratios of NH$_3$/HC$_3$N, NH$_3$/CCS, and N$_2$H$^+$/CCS are low in starless regions, while they are high in star-forming cores in the Orion A GMC \citep{oha14,tat14a}. They suggested that these molecules may be indicators of the chemical evolution in the Orion A GMC. \citet{tat14b} observed CCS and NH$_3$ emission toward a starless core in the Orion A GMC with the Very Large Array, and revealed that the CCS emission surrounds the NH$_3$ core. This configuration resembles that of the N$_2$H$^+$ and CCS distribution in the Taurus L1544 prestellar core showing the collapsing motion.
It is needed to establish whether these molecules can generally be indicators of the chemical evolution even in GMCs by investigating dense cores located in other GMCs.

We focus on one of near GMCs located in the southern hemisophere, Vela C molecular cloud complex.
The Vela molecular clouds or `Vela Molecular Ridge'  \citep{mur91} is one of closest GMCs within the Galactic plane, and consists of four components labelled A through D. 
The Vela C giant molecular cloud complex is the most massive component of the Vela region, and accompanies a bright HII region, RCW 36, associated with an early type star (spectral type O5-B0)  \citep{mas03}. The distance is 700 pc \citep{lis92}.  
\citet{yam99} mapped $^{12}$CO, $^{13}$CO, and C$^{18}$O ($J=1-0$) lines in Vela C, and detected molecular outflows. By $Herschel$ PACS and SPIRE observations at 70, 160, 250, 350, and 500 $\mu$m dust continuum emission, \citet{hil11} identified high column density filaments over the whole region, and showed differences in their column density and temperature probability distribution function (PDF) among regions. They revealed that the column density PDF of the `Centre-Ridge region', which contains a high-mass star, is flatter than other regions, and shows a high column density tail.

In this study, we investigate the chemical characteristics of molecular cloud cores in the Vela C giant molecular cloud complex through mapping observations of HC$_3$N and N$_2$H$^+$. The purpose of this study is to investigate whether HC$_3$N (a carbon-chain molecule)  and N$_2$H$^+$ (a N-bearing molecule) are indicators of  chemical evolution in a GMC.

\section{Observations}
\begin{figure}[htbp]
  \begin{center}
  \includegraphics[width=15cm]{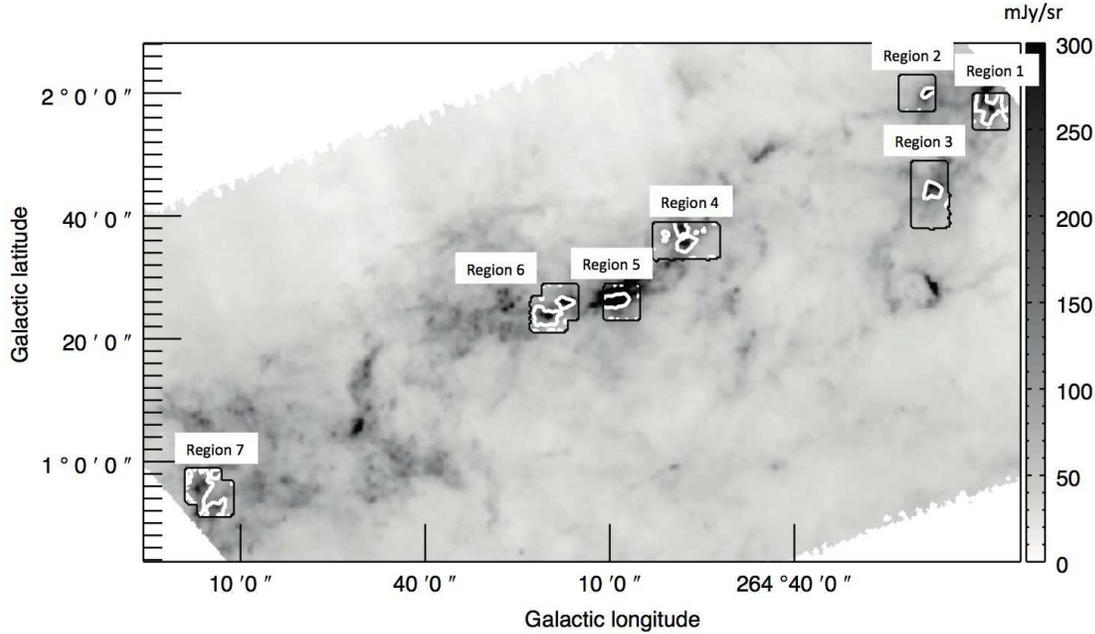}
  \end{center}
  \caption{The N$_2$H$^+$ $J=1-0$ velocity-integrated intensity maps in contours are imposed on the gray scale map of the $Herschel$ $SPIRE$ 500 $\mu$m dust emission. The velocity range of integration for the N$_2$H$^+$ line emission is from 3.5 to 9.0 km s$^{-1}$. The contours represent 0.4 K km s$^{-1}$.}
  \label{area}
\end{figure}

Observations were carried out  by the Mopra 22 m telescope located in Australia from 2014  May 10 to 25. We used the 3 mm MMIC receiver that can simultaneously record dual polarization data. We simultaneously observed HC$_3$N ($J=10-9$) at 90.978989 GHz and N$_2$H$^+$ ($J=1-0$) at 93.1737767 GHz.
The employed spectrometer was the Mopra Spectrometer (MOPS) digital filter
bank  in the zoom-band mode.
The spectral resolution was 33.57 kHz ($\sim0.11$ km s$^{-1}$).
 At 90 GHz, the half-power beam width (HPBW) and ``extended beam efficiency'' of the telescope are 42\arcsec and 50\%, respectively  \citep{lad05}.
 We observed source 81 of the Vela-D molecular cloud identified by \citet{mor12} once a day to check the stability of the intensity calibration. The absolute intensity accuracy was estimated to be less than 10\%.
We mapped 5$\arcmin$ $\times$ 5$\arcmin$ box areas in the on-the-fly (OTF) mode. We observed a total of 7 regions, which were selected from bright C$^{18}$O emission peaks.  The C$^{18}$O $(J=1-0)$ observations were previously carried out  by \citet{yam99} with 2$\arcmin$  resolution. Figure \ref{area} shows the locations of the observed areas superimposed on a $Herschel$ $SPIRE$ 500 $\mu$m map. The black boxes indicate OTF mapping area and white contours represent the N$_2$H$^+$ $J=1-0$ velocity-integrated intensity map. 
The obtained data were smoothed to a HPBW of 53$\arcsec$ with a 2D Gaussian function in order to improve the signal-to-noise ratio. The angular resolution is equivalent to 0.2 pc at a distance of 700 pc. It is worth noting that this resolution will identify objects that will form groups of stars. The typical rms noise level is about 0.08 K per channel. 
The telescope pointing was checked every 1.0 hour by observing the
SiO maser source L2 Pup (RA,DEC (J2000) $=$ 7$^h$13$^m$32$\fs$31, -44$^\circ$38$\arcmin$24$\farcs$1). The pointing accuracy was better than 10$\arcsec$.
The upper energy levels ($E_u/k$) for HC$_3$N ($J=10-9$) and N$_2$H$^+$ ($J=1-0$) transitions  are 24.01 and 4.47 K, respectively.   
The critical densities for these transitions are $3\times 10^5$ and $5\times 10^5$ cm$^{-3}$ at 20 K, respectively \citep{san12}.  However, \citet{shi15} suggested that N$_2$H$^+$ is excited at lower densities than the critical density, and the effective excitation densities may be lower than $5\times 10^5$ cm$^{-3}$.

The observed data were reduced using ``Livedata'' and ``Gridzilla'', which are the most typical data reduction packages for the Mopra telescope\footnote{http://www.atnf.csiro.au/computing/software/livedata/index.html}.
``Livedata'' fits and subtracts a linear baseline. The output of ``Livedata'' is recorded in single-dish fits format (sdfits).
``Gridzilla''  uses this output to regrid data onto a data cube using a Gaussian smoothing. The final data cubes were smoothed to 15$\arcsec$ grid size.

\section{Results and Disccusion}
\subsection{Integrated Intensity maps}
\begin{figure}[htbp]
  \begin{center}
  \includegraphics[width=12cm]{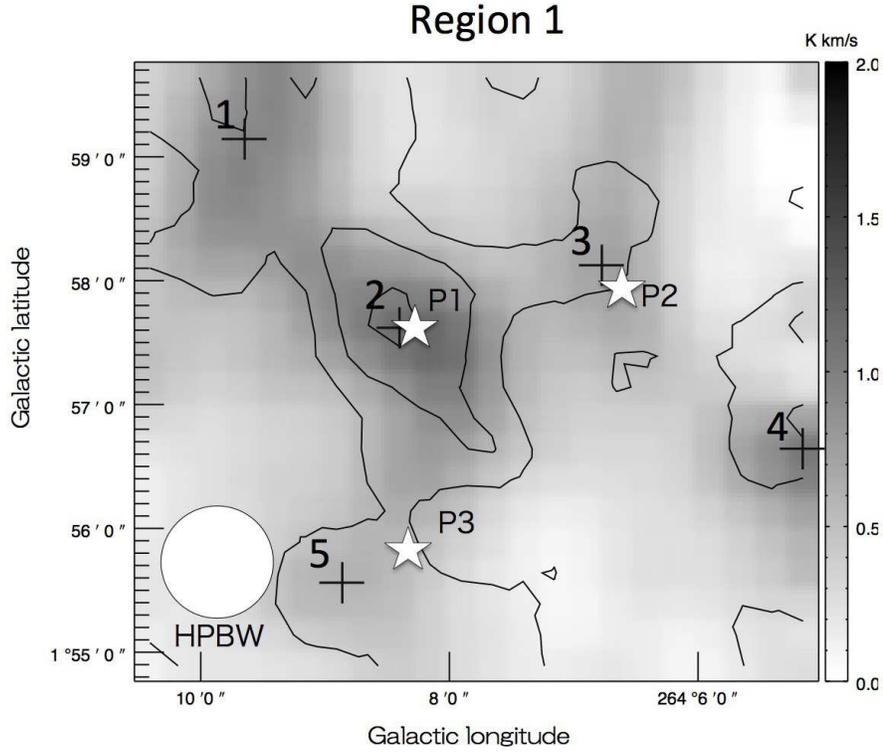}
  \end{center}
  \caption{The HC$_3$N $J=10-9$ velocity-integrated intensity  map is superimposed on the gray scale map of the integrated intensity of the main hyperfine component group N$_2$H$^+$ $J = 1-0$, $F_1 = 2-1$. The integrated velocity range for both molecular lines is from 5.0 to 9.0 km s$^{-1}$. The lowest contour and the contour step are 3 $\sigma$. The 1 $\sigma$ noise level for the contour is 0.05 K km s$^{-1}$.  The white star signs represent the locations of the protostars in \citet{gia12}. The plus signs represent the locations of the N$_2$H$^+$ and HC$_3$N intensity peaks.}
  \label{region1}
\end{figure}

\begin{figure}[htbp]
  \begin{center}
  \includegraphics[width=12cm]{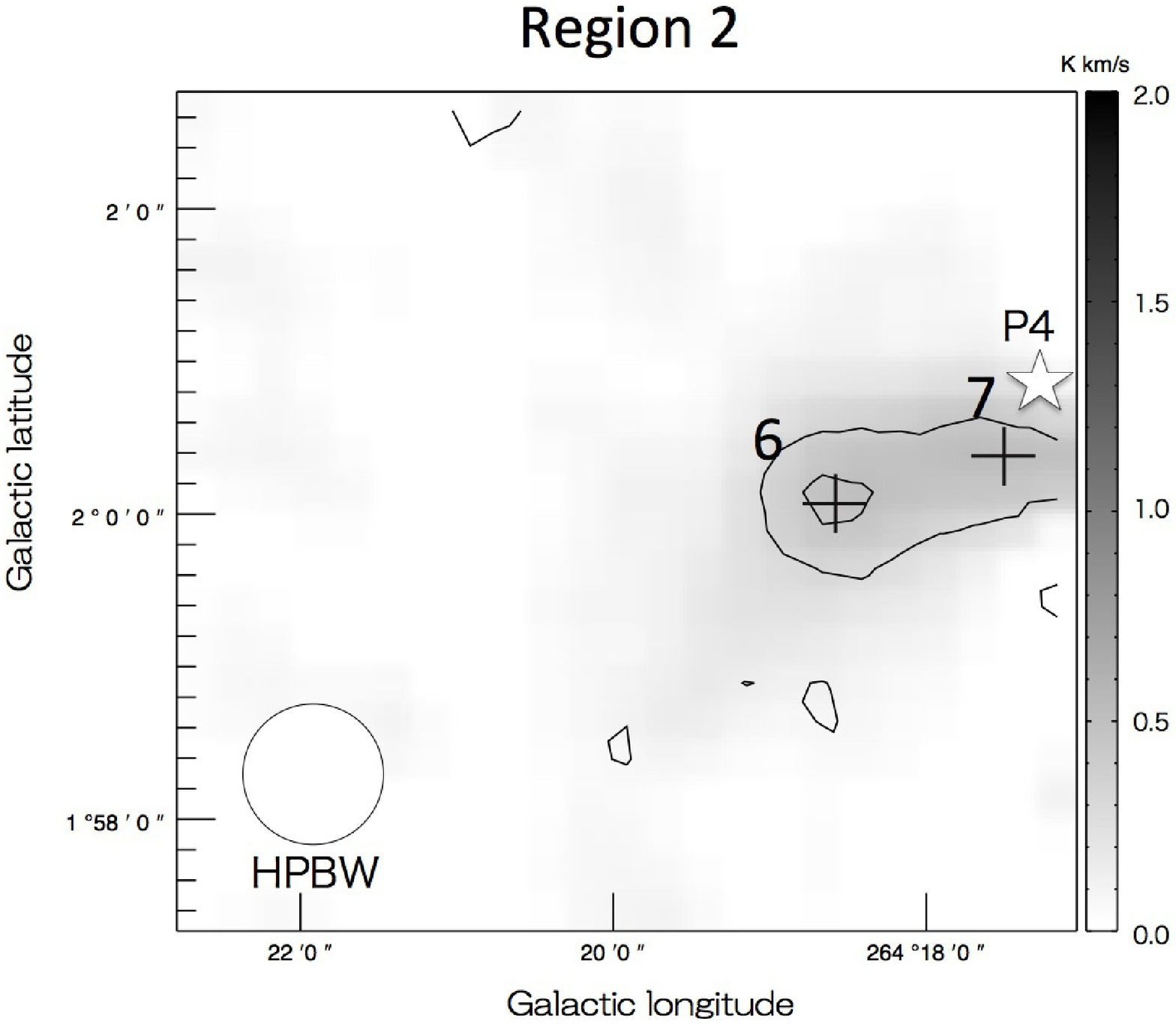}
  \end{center}
  \caption{The same as figure \ref{region1} but for region 2. The integrated velocity range for the molecular line emission is from 6.0 to 8.0 km s$^{-1}$. The lowest contour and the contour step are 3 $\sigma$. The 1 $\sigma$ noise level is 0.04 K km s$^{-1}$.  }
  \label{region2}
\end{figure}

\begin{figure}[htbp]
  \begin{center}
  \includegraphics[width=12cm]{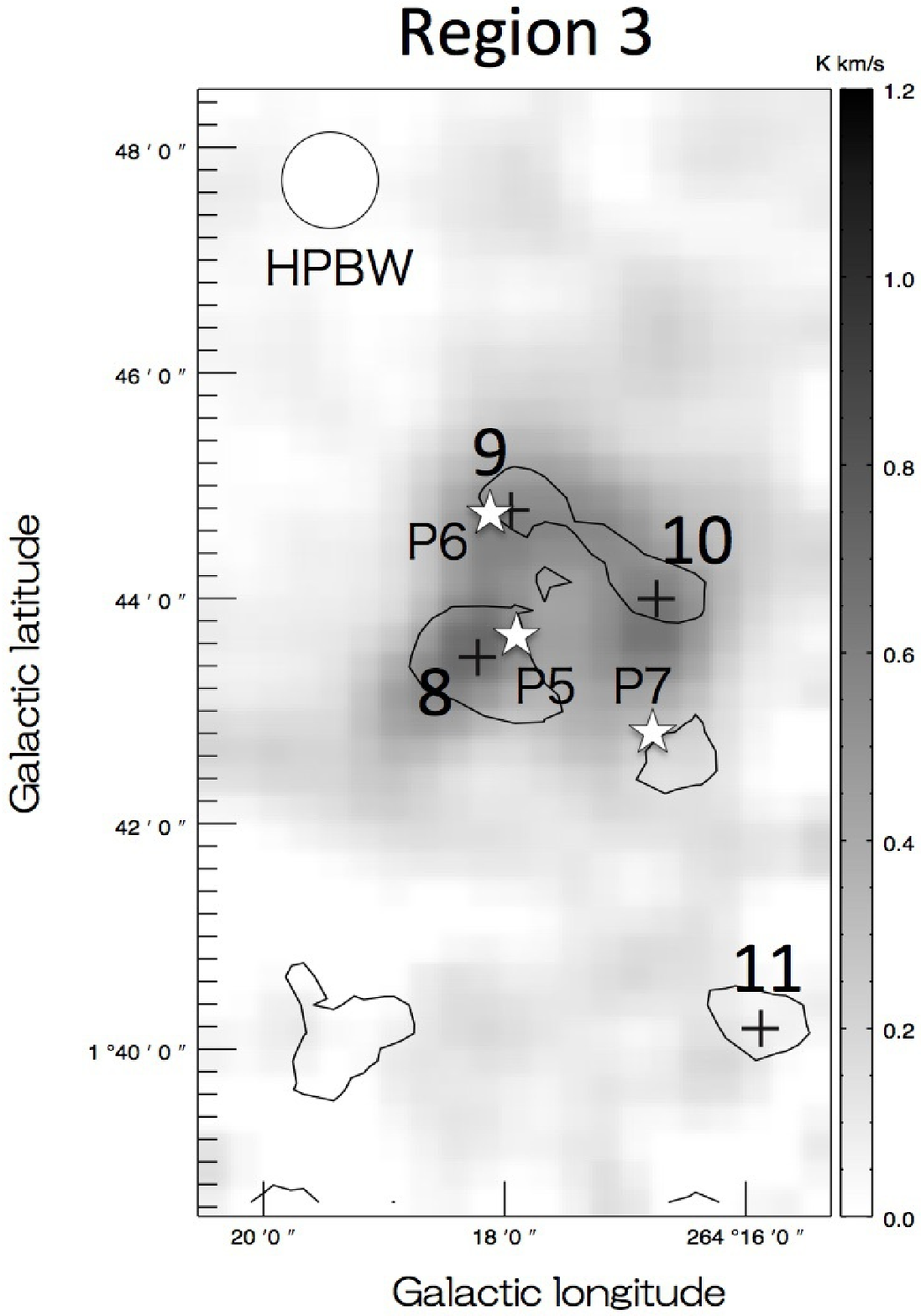}
  \end{center}
  \caption{The same as figure \ref{region1} but for region 3. The integrated velocity range for the molecular line emission is from 6.0 to 8.0 km s$^{-1}$. The lowest contour and the contour step are 3 $\sigma$. The 1 $\sigma$ noise level is 0.04 K km s$^{-1}$.}
  \label{region3}
\end{figure}

\begin{figure}[htbp]
  \begin{center}
  \includegraphics[width=12cm]{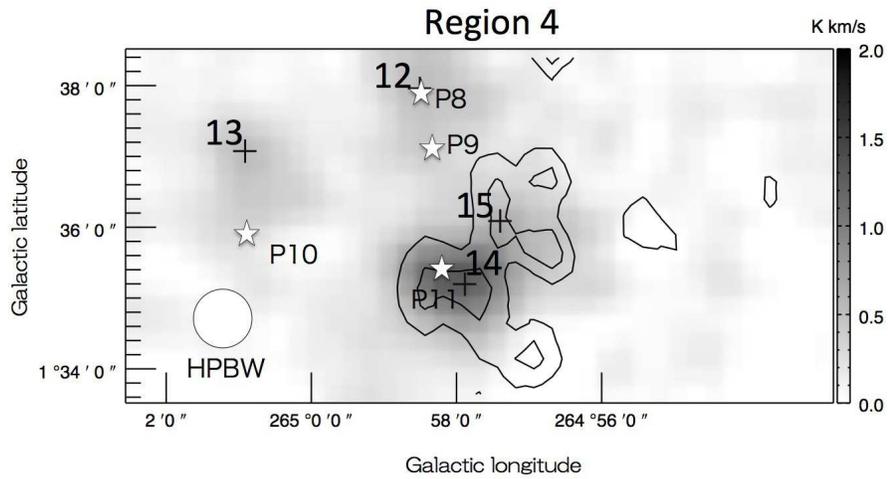}
  \end{center}
  \caption{The same as figure \ref{region1} but for region 4. The integrated velocity range for the molecular line emission is from 5.0 to 7.5 km s$^{-1}$ . The lowest contour is 3 $\sigma$, and the contour step is 2 $\sigma$. The 1 $\sigma$ noise level is 0.05 K km s$^{-1}$.}
  \label{region4}
\end{figure}

\begin{figure}[htbp]
  \begin{center}
  \includegraphics[width=12cm]{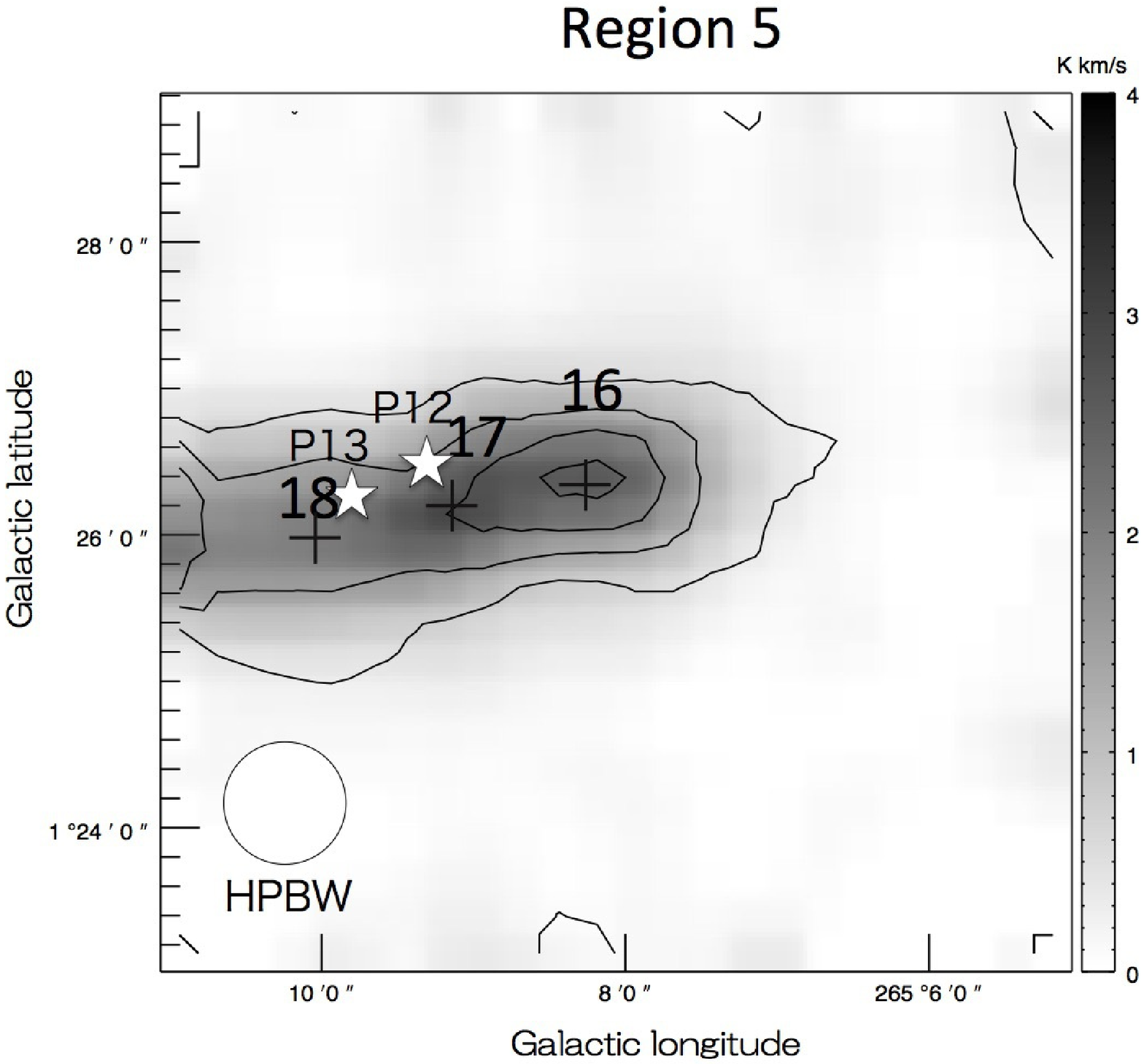}
  \end{center}
  \caption{The same as figure \ref{region1} but for region 5. The integrated velocity range for the molecular line emission is from 6.0 to 9.0 km s$^{-1}$. The lowest contour  and the contour step are 3 $\sigma$. The 1 $\sigma$ noise level is 0.06 K km s$^{-1}$.}
  \label{region5}
\end{figure}

\begin{figure}[htbp]
  \begin{center}
  \includegraphics[width=12cm]{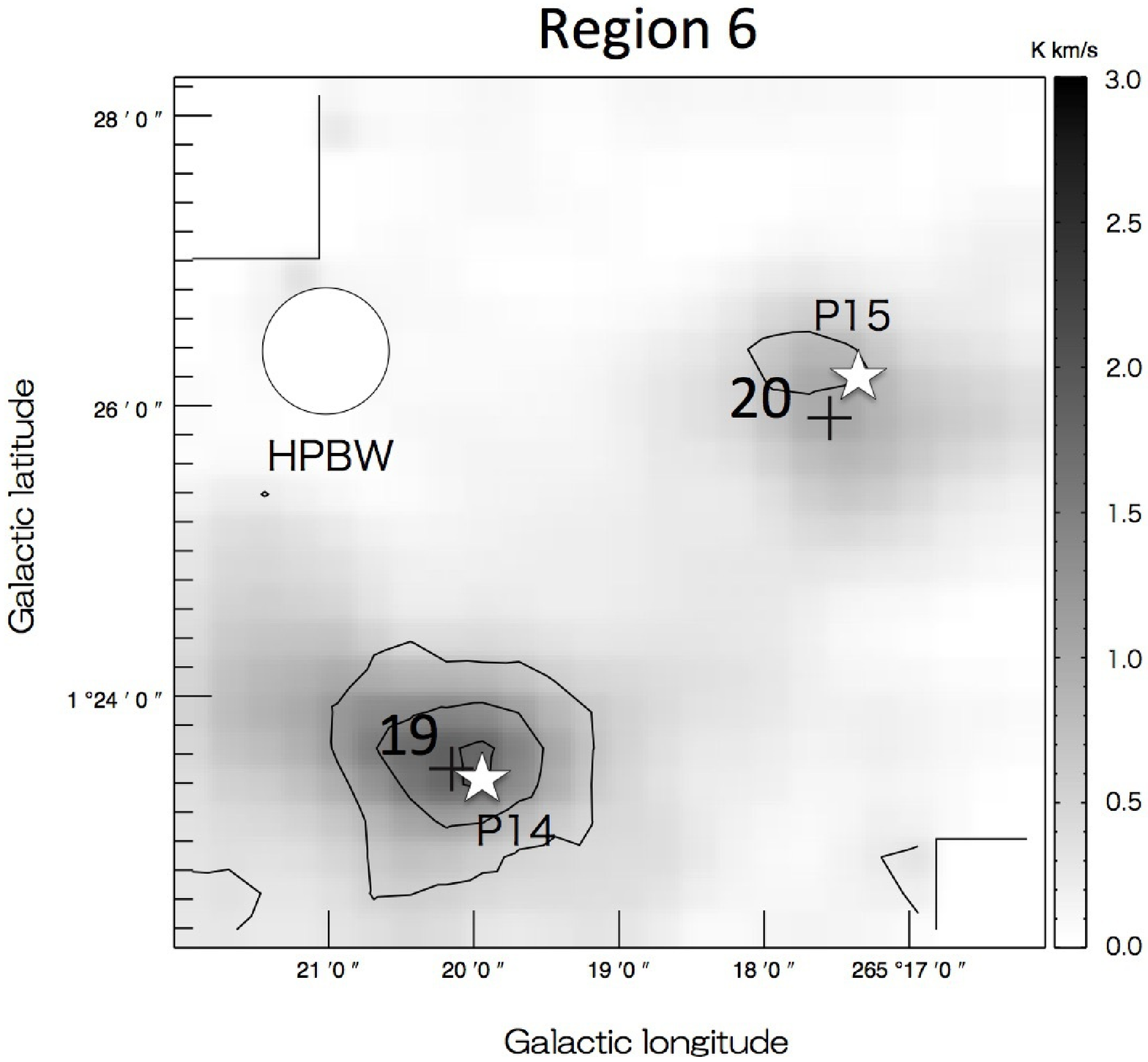}
  \end{center}
  \caption{The same as figure \ref{region1} but for region 6. The integrated velocity range for the molecular line emission is  from 6.0 to 9.0 km s$^{-1}$. The lowest contour and the contour step are 3 $\sigma$. The 1 $\sigma$ noise level is 0.05 K km s$^{-1}$.}
  \label{region6}
\end{figure}

\begin{figure}[htbp]
  \begin{center}
  \includegraphics[width=12cm]{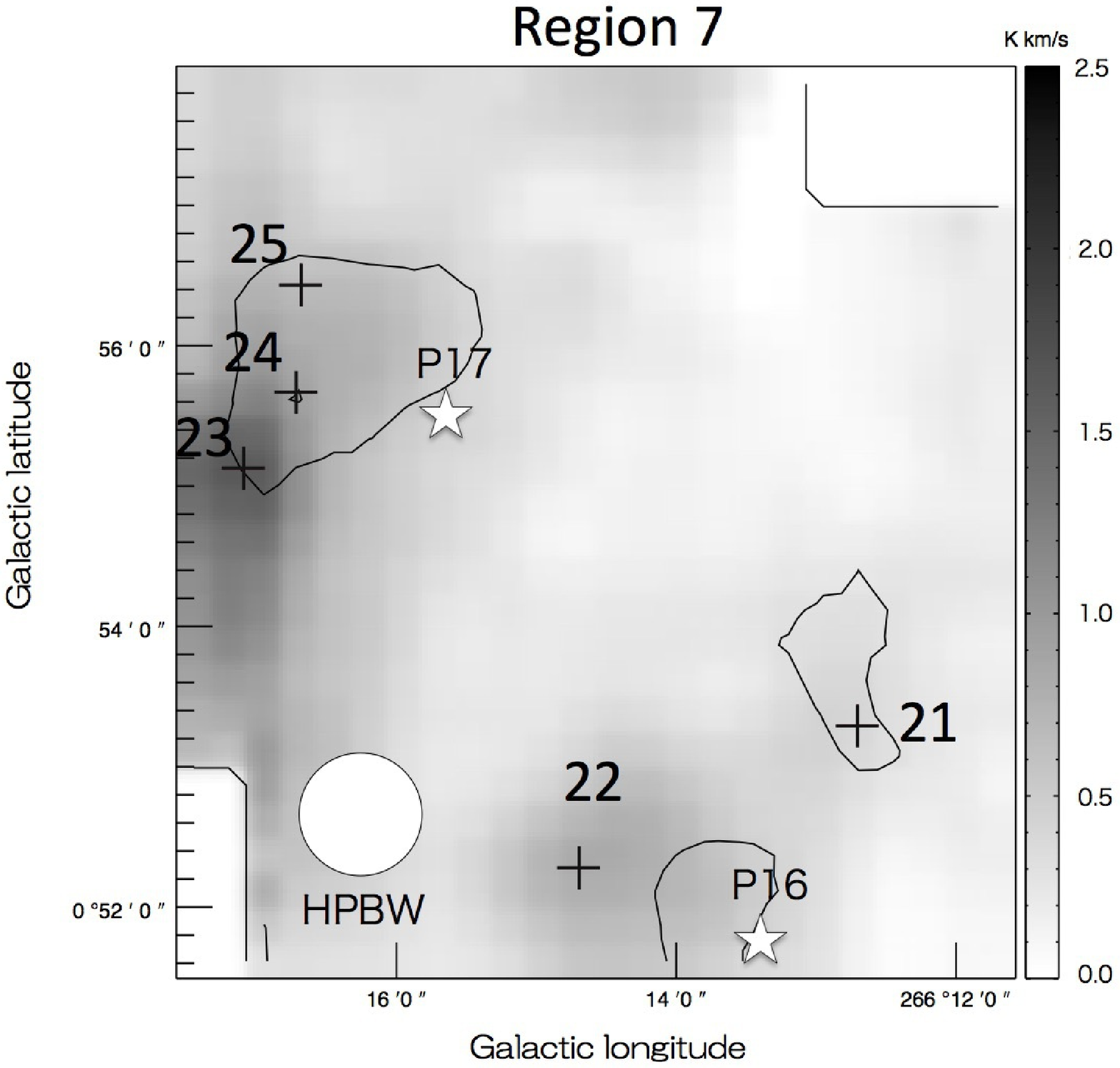}
  \end{center}
  \caption{The same as figure \ref{region1} but for region 7. The integrated velocity range for the molecular line emission is from 3.5 to 6.0 km s$^{-1}$. The lowest contour  and the contour step are 3 $\sigma$. The 1 $\sigma$ noise level is 0.07 K km s$^{-1}$.}
  \label{region7}
\end{figure}

Figures \ref{region1} through \ref{region7} show the integrated intensity maps in $T_{\rm A}^*$ scale.  The HC$_3$N $J=10-9$ contour maps superimposed on the N$_2$H$^+$ $J=1-0$ gray-scale images. We plot the location of  protostars identified by thermal emission at 5 wavebands from 70 to 500 $\mu$m using $PACS$ and $SPIRE$ cameras on board the $Herschel$ $Space$ $Observatory$ (\citet{gia12}; T. Giannini, private communication for the protostar coordinates).   We also plot protostar candidates identified by the near-infrared ($J$, $H$, $K_S$) survey carried out by \citet{bab06} and AKARI-FIS Point Source Catalogue, which are Class 0 protostellar candidates \citep{sun09}.
 We identified 25 local peaks in the velocity-integrated intensity maps of the HC$_3$N $J=10-9$ and/or N$_2$H$^+$ $J=1-0$ emission in these 7 regions. The criteria for identification are more than five sigma above the noise level for N$_2$H$^+$ and/or three sigma above the noise level for HC$_3$N. If the HC$_3$N peak is within the HPBW of the N$_2$H$^+$ peak, we use the HC$_3$N peak position.  The identification of N$_2$H$^+$ emission peaks is more strict so that we can perform hyperfine fitting to satellite components. In Table \ref{tab:hc3n}, we list the positions of the detected HC$_3$N $J=10-9$ and/or N$_2$H$^+$ $J=1-0$ local peaks in the velocity-integrated intensity map and the line parameters of the HC$_3$N spectra measured toward these positions. The peak intensities and linewidths are obtained through Gaussian fitting. If  local intensity peaks coincide with the protostars within 53$\arcsec$, we define the peaks as star-forming cores. If local intensity peaks do not coincide with the protostars, we define the peaks as starless cores.  A total of 12 peak positions were identified as star-forming. 
We assume that both lines trace sufficiently dense gas, and the starless cores identified in these lines will eventually form protostars.
We found that the distributions of  the HC$_3$N and N$_2$H$^+$ velocity-integrated intensities are similar to each other, suggesting that these molecular lines trace the same dense cores.   Taking into account the fact that  the scale of different distributions between HC$_3$N and N$_2$H$^+$ occurs at 0.1 pc scales in OMC-3 region (\cite{tat08}),  less prominent differences of intensity distributions  would be due to our larger spatial resolution of 0.19 pc. 

\begin{longtable}[htbp]{ccccccc}

 \caption{ HC$_3$N and N$_2$H$^+$ intensity peaks, coordinates, and HC$_3$N parameters. }\label{tab:hc3n}
  \hline 

Peak Position	&	$l$	&	$b$	&	$T_{A}^*$		&	$v_{\rm LSR}$(HC$_3$N)	&	$dv$(HC$_3$N)	& Protostar		 \\
name	&	$\degree$	&	$\degree$ 	&	(K)	&	(km s$^{-1}$)	&	(km s$^{-1}$)	 & 	\\
	\endfirsthead
\hline
\endhead
  \hline
\endfoot
  \hline
\endlastfoot
  \hline

1	&	264.161	&	1.986	&		&		&		&		\\
2	&	264.140	&	1.960	&	0.41$\pm$0.06	&	6.6$\pm$0.1	&	1.00$\pm$0.18	&	P1	\\
3	&	264.113	&	1.969	&	0.20$\pm$0.07	&	6.8$\pm$0.1	&	0.75$\pm$0.31	&	P2	\\
4	&	264.086	&	1.944	&	0.22$\pm$0.05	&	6.5$\pm$0.3	&	1.10$\pm$0.59	&		\\
5	&	264.148	&	1.926	&	0.26$\pm$0.06	&	6.9$\pm$0.1	&	0.66$\pm$0.15	&	P3	\\
6	&	264.31	&	1.733	&	0.30$\pm$0.07	&	7.1$\pm$0.1	&	0.83$\pm$0.21	&		\\
7	&	264.292	&	2.006	&	0.16$\pm$0.07	&	7.2$\pm$0.2	&	0.70$\pm$0.35	&	P4	\\
8	&	264.304	&	1.725	&	0.23$\pm$0.06	&	6.6$\pm$0.1	&	0.99$\pm$0.31	&	P5	\\
9	&	264.299	&	1.746	&		&		&		&	P6	\\
10	&	264.279	&	1.733	&	0.24$\pm$0.09	&	6.8$\pm$0.1	&	0.37$\pm$0.21	&		\\
11	&	264.265	&	1.67	&	0.21$\pm$0.07	&	6.9$\pm$0.1	&	0.77$\pm$0.30	&		\\
12	&	264.975	&	1.633	&		&		&		&	P8	\\
13	&	265.015	&	1.618	&	0.07$\pm$0.06	&	7.0$\pm$0.4	&	1.02$\pm$1.00	&		\\
14	&	264.965	&	1.587	&	0.29$\pm$0.06	&	6.4$\pm$0.1	&	0.99$\pm$0.24	&	P11	\\
15	&	264.957	&	1.601	&	0.26$\pm$0.06	&	6.4$\pm$0.1	&	1.00$\pm$0.27	&		\\
16	&	265.138	&	1.439	&	0.35$\pm$0.05	&	8.3$\pm$0.1	&	1.37$\pm$0.71	&		\\
	&	265.138	&	1.439	&	0.24$\pm$0.03	&	6.6$\pm$0.2	&	2.29$\pm$0.34	&		\\
17	&	265.152	&	1.437	&	0.32$\pm$0.04	&	7.1$\pm$0.1	&	2.05$\pm$0.32	&	P12	\\
18	&	265.167	&	1.433	&	0.29$\pm$0.04	&	7.2$\pm$0.2	&	2.14$\pm$0.34	&	P13	\\
19	&	265.336	&	1.392	&	0.33$\pm$0.06	&	6.5$\pm$0.1	&	1.18$\pm$0.23	&	P14	\\
20	&	265.292	&	1.432	&	0.13$\pm$0.09	&	6.8$\pm$0.1	&	0.85$\pm$0.24	&	P15	\\
21	&	266.212	&	0.888	&	0.36$\pm$0.08	&	4.9$\pm$0.1	&	0.60$\pm$0.14	&		\\
22	&	266.245	&	0.871	&	0.23$\pm$0.08	&	5.1$\pm$0.1	&	0.67$\pm$0.26	&		\\
23	&	266.285	&	0.918	&		&		&		&		\\
24	&	266.279	&	0.928	&	0.29$\pm$0.06	&	4.5$\pm$0.1	&	1.24$\pm$0.27	&		\\
25	&	266.278	&	0.941	&	0.22$\pm$0.05	&	4.8$\pm$0.2	&	1.40$\pm$0.37	&		\\
\end{longtable}

\subsection{Hyperfine fitting of N$_2$H$^+$ and Column density}
\begin{figure}[htbp]
  \begin{center}
  \includegraphics[width=12cm]{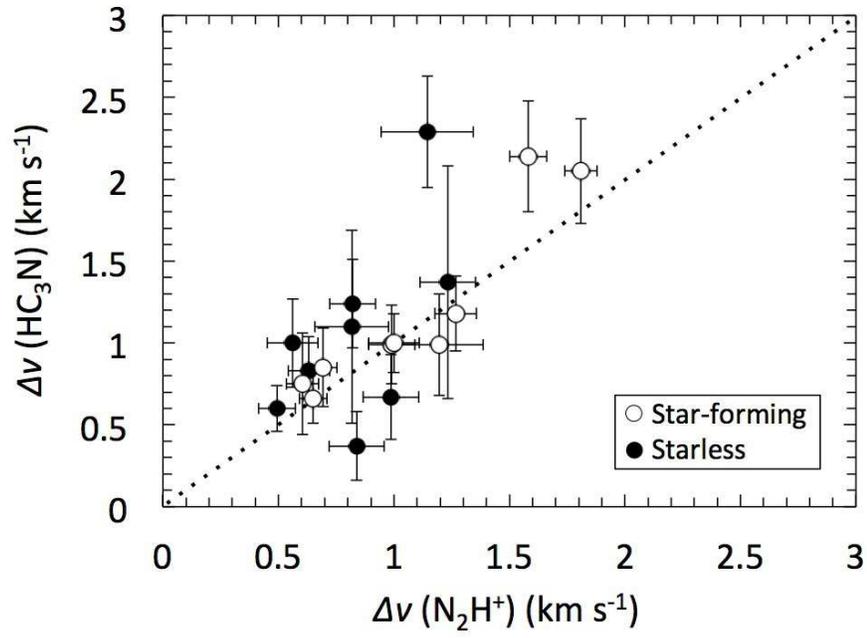}
  \end{center}
  \caption{The HC$_3$N linewidth against that the N$_2$H$^+$ linewidth. The dotted line represents unity.}
  \label{fig2}
\end{figure}

\begin{figure}[htbp]
  \begin{center}
  \includegraphics[width=18cm]{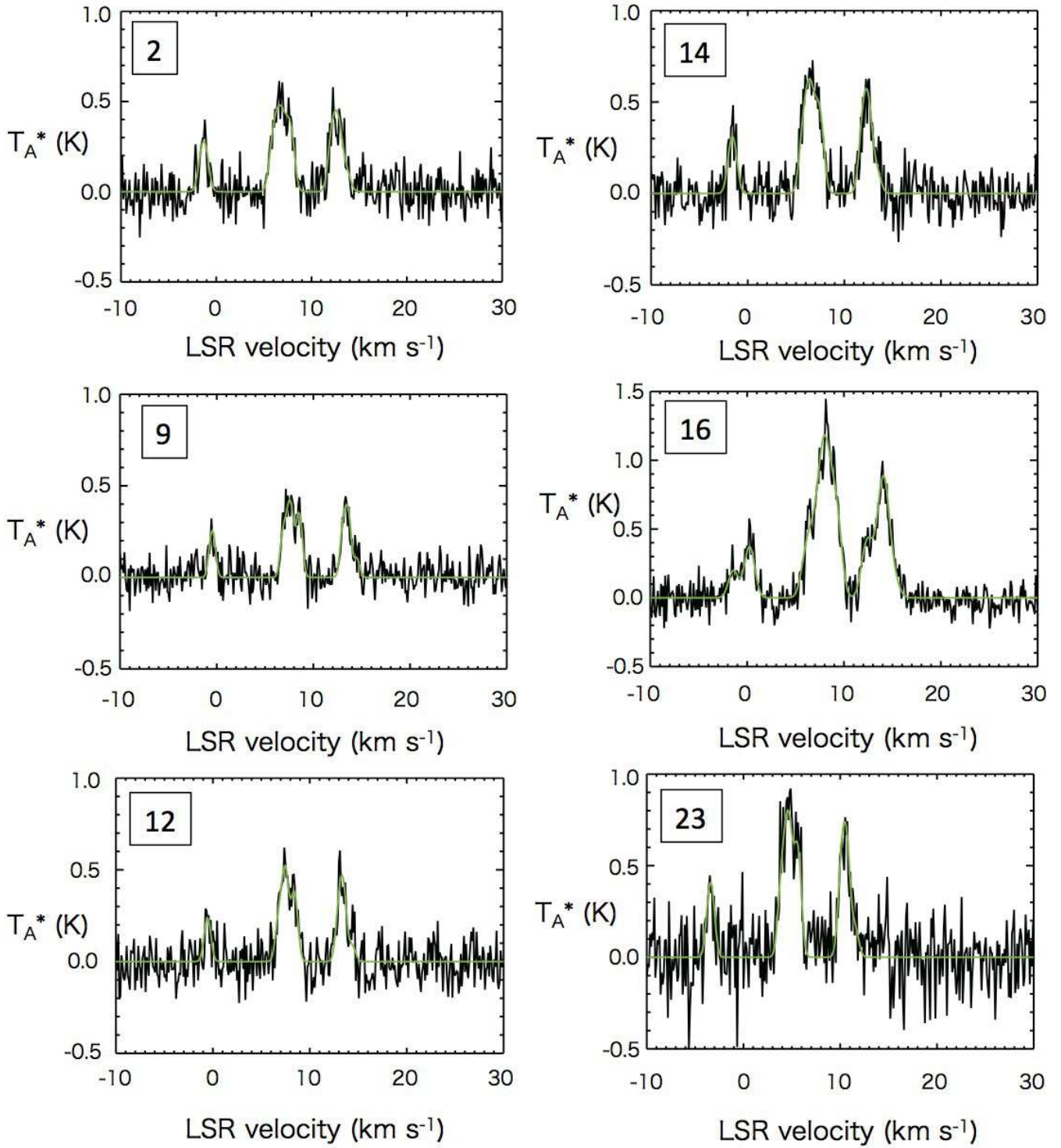}
  \end{center}
  \caption{Hyperfine-fitting results for the N$_2$H$^+$ ($J=1-0$) spectra for the six intensity peaks. 
  }
  \label{n2hfit}
\end{figure}

We fitted a hyperfine component model to the N$_2$H$^+$ spectra, and derived the optical
depth, LSR velocity, linewidth, and excitation temperature assuming a uniform excitation temperature in the N$_2$H$^+$ hyperfine components. We also assume that the N$_2$H$^+$ emission fills the whole beam, that is, the beam filling factor is equal to unity. Figure \ref{fig2} plots the linewidths of HC$_3$N against those of N$_2$H$^+$. The dashed line delineates the case that they are equal. We found a good correlation between them, suggesting that the molecular emission likely comes from  the same volume. This is supported by the fact that the centroid velocities of the N$_2$H$^+$ and HC$_3$N lines are consistent with each other (see also Table 1 and 2).  The intrinsic line strengths of the
hyperfine components are adopted from \citet{tin00}. 
The optical depth $\tau_{\rm tot}$ is the sum of the optical depths of all the hyperfine components.
At peak 16, we identified two velocity components and we fit a two-velocity-component hyperfine model. Figure \ref{n2hfit} shows examples of hyperfine-fitting results at the peak positions of the cores.
The column density
is calculated by assuming local thermodynamic equilibrium (LTE). The formulation can be
found, for example, in Equation (96) and (97) of \citet{man15}.  The range of the optical depths and the excitation temperatures are: $\tau_{\rm tot} = 0.9 - 9.6$ and $T_{\rm ex} = 3.3 - 10.4$ K (see also Table \ref{tab:column}). Even in the most prominent hyperfine line ($F_1=2-1$, $F=3-2$), the optical thickness is 0.259 $\tau_{\rm tot}$ (N$_2$H$^+$) \citep{tin00}.  Therefore, each N$_2$H$^+$ hyperfine emission line is optically thin in general because the median value of $\tau_{\rm tot}$(N$_2$H$^+$) was estimated to be $\sim$2.4.  The 1 $\sigma$ error of the N$_2$H$^+$ column density was estimated by $\Delta \tau_{\rm tot}$ of the hyperfine fitting results. However, if we use the lower limit of the $\tau_{\rm tot}-\Delta \tau_{\rm tot}$, the $T_{\rm ex}$(N$_2$H$^+$) can be too high to accept, that is, $T_{\rm ex}$(N$_2$H$^+$) is higher than $T_{\rm ex}$(CO) $\sim$ $10-20$ K from \citet{yam99}. In these cases, we use the symmetric error bars in the log scale by adopting the upper limit $\tau_{\rm tot}+\Delta \tau_{\rm tot}$.

For HC$_3$N, the optical depth has been derived by using multitransitional observations ($J=4-3$, $J=10-9$, $J=12-11$, and $J=16-15$) toward several GMCs including Orion A GMC by \citet{ber96}, and the opacity was found to be $<$1 at all positions. The HC$_3$N column density is calculated by assuming LTE and optically thin emission. 
The formulation can be found in Equation (13) of \citet{san12}, for example. The excitation temperature $T_{\rm ex}$(HC$_3$N) is assumed to be equal to that for N$_2$H$^+$. This assumption has some uncertainties, but the same excitation will be a best guess because both lines are thought to be optically thin and would be only subthermally excited. We confirmed that the excitation temperature of N$_2$H$^+$ does not change between the star-forming and starless peaks suggesting that the excitation condition of the dense cores would not be changed.  When hyperfine fitting is not successful due to low signal to noise ratios, we adopt $T_{\rm ex}$(HC$_3$N) = 5 K, which is the average value of $T_{\rm ex}$(N$_2$H$^+$).  In the case of non-detection of these lines, we estimate the upper limit of the column densities from 3 sigma noise level.  
Table \ref{tab:column} summarizes the derived column densities.

\begin{longtable}[htbp]{cllllll}

 \caption{ N$_2$H$^+$ hyperfine fitting results and column densities. }\label{tab:column}
  \hline 

Peak Position	&	$\tau_{\rm tot}$(N$_2$H$^+$)	&	$v_{\rm LSR}$ (N$_2$H$^+$)	&	$dv$(N$_2$H$^+$)		&	$T_{\rm ex}$	&	$N$(N$_2$H$^+$)		&	$N$(HC$_3$N) \\
	name		&							&	(km s$^{-1}$)				&	(km s$^{-1}$)	&	(K)		&	(cm$^{-2}$)	 	&	(cm$^{-2}$)	\\
	\endfirsthead
\hline
\endhead
  \hline
\endfoot
  \hline
\endlastfoot
  \hline

1	&	1.2$\pm1.1$	&	7.40$\pm0.06$	&	1.70$\pm0.21$	&	4.7$\pm1.5$	&	(5.7$^{+5.3}_{-2.7}$)E+12	&	$<$7.50E+12	\\
2	&	6.2$\pm2.2$	&	6.68$\pm0.04$&	1.00$\pm0.11$	&	3.8$\pm0.2$	&	(1.3$^{+4.8}_{-3.5}$)E+13	&	(1.2$\pm0.3$)E+13	\\
3	&	3.5$\pm2.9$	&	6.78$\pm0.03$	&	0.60$\pm0.07$	&	4.2$\pm0.8$	&	(5.0$^{+4.2}_{-2.3}$)E+12	&	(3.4$\pm1.8$)E+12	\\
4	&	5.2$\pm$4.2	&	7.04$\pm$0.06	&	0.82$\pm$0.16	&	3.9$\pm$0.5	&	(9.1$^{+7.3}_{-4.1}$)E+12	&	(6.6$\pm3.8$)E+12	\\
5	&	5.6$\pm$2.6	&	6.96$\pm$0.03	&	0.65$\pm$0.06	&	3.6$\pm$0.2	&	(7.0$^{+3.3}_{-2.2}$)E+12	&	(7.3$\pm1.9$)E+12	\\
6	&	3.9$\pm3.6$	&	7.22$\pm0.04$&	0.63$\pm0.09$	&	3.9$\pm0.7$	&	(5.2$^{+5.0}_{-2.6}$)E+12	&	(6.7$\pm2.3$)E+12	\\
7	&				&				&				&	5			&	$<$ 2.2E+12		&	(1.9$\pm1.3$)E+12	\\
8	&	3.6$\pm2.4$	&	6.88$\pm0.06$	&	1.20$\pm0.19$	&	3.7$\pm0.4$	&	(8.6$^{+5.7}_{-3.4}$)E+12	&	(7.4$\pm3.1$)E+12	\\
9	&	6.0$\pm2.6$	&	7.54$\pm0.03$	&	0.70$\pm0.07$	&	3.8$\pm0.2$	&	(8.6$^{+3.5}_{-2.5}$)E+12	&	$<$5.0E+12	\\
10	&	3.0$\pm2.6$	&	6.84$\pm0.05$	&	0.84$\pm0.12$	&	4.1$\pm0.8$	&	(5.8$^{+5.2}_{-2.7}$)E+12	&	(2.1$\pm1.4$)E+12	\\
11	&				&				&				&	5			&	$<$2.4E+12		&	(2.7$\pm1.4$)E+12	\\
12	&	1.7$\pm1.6$	&	7.44$\pm0.03$	&	0.80$\pm0.09$	&	5.4$\pm1.7$	&	(4.8$^{+4.7}_{-2.4}$)E+12	&	$<$2.9E+12	\\
13	&				&				&				&	5			&	$<$ 3.1E+12		&	(1.2$\pm1.2$)E+12	\\
14	&	3.8$\pm1.7$	&	6.41$\pm0.03$	&	0.99$\pm0.10$	&	4.5$\pm0.5$	&	(1.0$^{+4.4}_{-3.1}$)E+13	&	(5.6$\pm1.8$)E+12	\\
15	&	9.6$\pm7.5$	&	6.63$\pm0.05$	&	0.56$\pm0.11$	&	3.4$\pm0.2$	&	(9.7$^{+7.6}_{-4.3}$)E+12	&	(1.1$\pm0.4$)E+13	\\
16	&	1.0$\pm0.8$	&	8.28$\pm0.05$	&	1.23$\pm0.12$	&	9.2$\pm4.5$	&	(1.0$^{+0.8}_{-0.5}$)E+13	&	(5.9$\pm3.2$)E+12	\\
	&	1.9$\pm1.7$	&	6.61$\pm0.10$	&	1.14$\pm0.20$	&	4.8$\pm1.5$	&	(6.2$^{+5.8}_{-3.0}$)E+12	&	(9.8$\pm1.9$)E+12	\\
17	&	0.9$\pm0.4$	&	7.12$\pm0.02$	&	1.81$\pm0.07$	&	10.4$\pm3.1$	&	(1.7$^{+8.2}_{-5.5}$)E+13	&	(8.1$\pm1.7$)E+12	\\
18	&	2.1$\pm0.6$	&	7.23$\pm0.03$	&	1.58$\pm0.08$	&	5.9$\pm0.7$	&	(1.3$^{+0.4}_{-0.3}$)E+13	&	(8.9$\pm1.9$)E+12	\\
19	&	1.7$\pm0.9$	&	6.46$\pm0.03$	&	1.27$\pm0.09$	&	7.0$\pm1.8$	&	(1.1$^{+0.6}_{-0.4}$)E+13	&	(5.1$\pm1.4$)E+12	\\
20	&	3.0$\pm2.0$	&	6.85$\pm0.03$	&	0.69$\pm0.06$	&	4.9$\pm1.0$	&	(6.3$^{+4.2}_{-2.5}$)E+12	&	(1.9$\pm1.4$)E+12	\\
21	&	9.1$\pm6.4$	&	4.88$\pm0.04$	&	0.49$\pm0.08$	&	3.3$\pm0.2$	&	(7.7$^{+5.4}_{-3.2}$)E+12	&	(1.1$\pm0.4$)E+13	\\
22	&	1.4$\pm1.4$	&	4.98$\pm0.04$	&	0.99$\pm0.12$	&	5.3$\pm1.4$	&	(4.7$^{+4.6}_{-2.3}$)E+12	&	(5.6$\pm1.2$)E+12	\\
23	&	3.7$\pm2.5$	&	4.60$\pm0.04$	&	0.87$\pm0.12$	&	5.2$\pm1.1$	&	(1.0$^{+0.7}_{-0.4}$)E+13	&	$<$3.4E+12	\\
24	&	2.8$\pm2.3$	&	4.59$\pm0.04$	&	0.82$\pm0.10$	&	4.6$\pm1.1$	&	(6.2$^{+5.2}_{-2.8}$)E+12	&	(6.9$\pm1.9$)E+12	\\
25	&				&				&				&	5			&	$<$ 4.3E+12		&	(5.2$\pm1.8$)E+12	\\
\end{longtable}

\subsection{Integrated intensity ratio and abundance ratio}

\begin{figure}[htbp]
  \begin{center}
  \includegraphics[width=11cm]{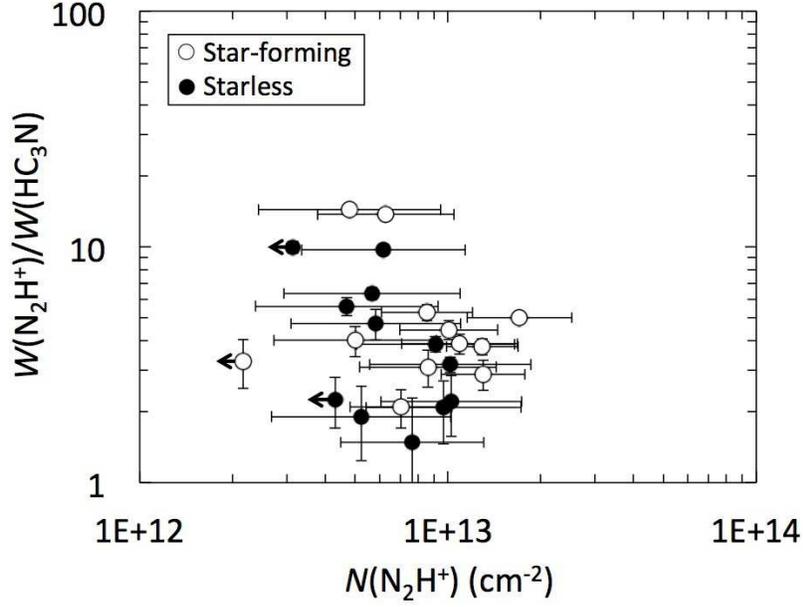}
  \end{center}
  \caption{The integrated intensity ratio of N$_2$H$^+$ to HC$_3$N  against the N$_2$H$^+$ column density.  The vertical error bar represents the 1 $\sigma$ noise level of integrated intensity ratio of N$_2$H$^+$ to HC$_3$N. The horizontal error bar represents the 1 $\sigma$ error corresponding to the N$_2$H$^+$ $J =1-0$ hyperfine line fitting.}
  \label{wratio}
\end{figure}

We show the ratio of the velocity-integrated intensity of the N$_2$H$^+$ main hyperfine group $J=1-0$, $F_1=2-1$ to that of the HC$_3$N $J=10-9$ emission against the N$_2$H$^+$ column density in Figure \ref{wratio}.
The integrated intensity ratios of N$_2$H$^+$/HC$_3$N are always greater than unity. It is also found that the star-forming peaks have high ratios compared with those of starless peaks, suggesting that the N$_2$H$^+$/HC$_3$N integrated intensity ratio increases with time as star formation evolves. However,  the substantial overlap between the star-forming and starless peaks can be seen.  This unclear boundary will be discussed in the following section.

\begin{figure}[htbp]
  \begin{center}
  \includegraphics[width=12cm]{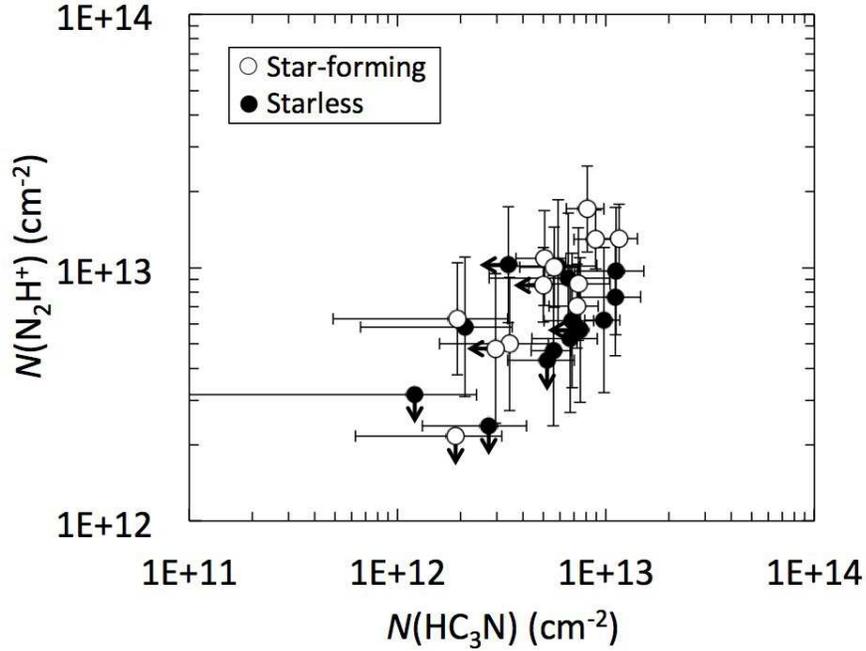}
  \end{center}
  \caption{The N$_2$H$^+$ versus  HC$_3$N column density. The vertical error bar represents the 1 $\sigma$ error corresponding to the N$_2$H$^+$ $J =1-0$ hyperfine line fitting. The horizontal error bar represents the 1 $\sigma$ error in the HC$_3$N assuming  that the excitation temperature $T_{\rm ex}$(HC$_3$N) is equal to that for N$_2$H$^+$.}
  \label{column}
\end{figure}

\begin{figure}[htbp]
  \begin{center}
  \includegraphics[width=12cm]{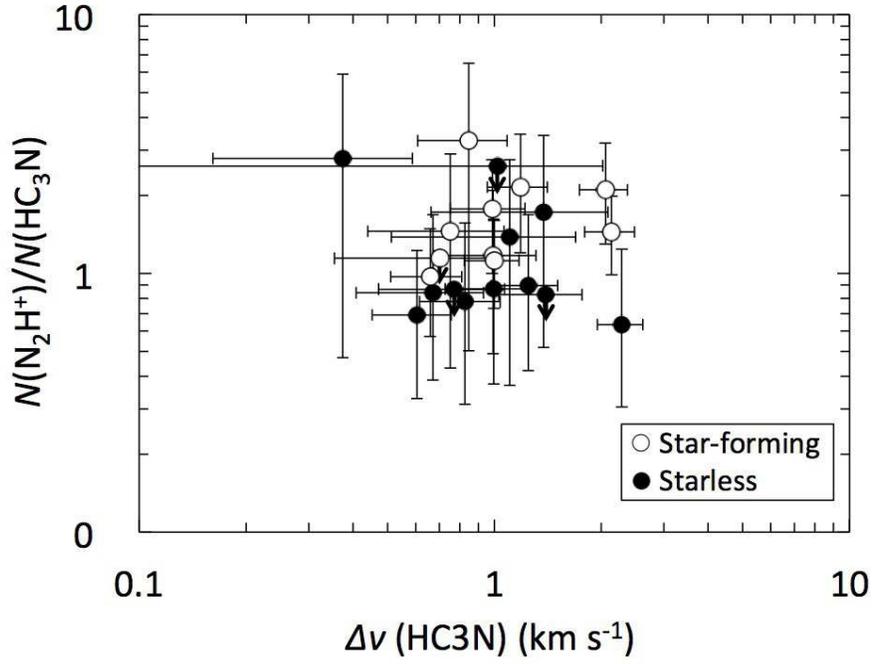}
  \end{center}
  \caption{The column density ratio of N$_2$H$^+$/HC$_3$N against the linewidth of the HC$_3$N emission. }
  \label{fig1}
\end{figure}

In molecular dense cores located in the Orion A GMC, it has been shown that the column density ratios of N$_2$H$^+$/CCS, NH$_3$/CCS, and NH$_3$/HC$_3$N are high in star-forming regions, while these are low in starless region (\cite{oha14,tat14a}).
These results suggest that the carbon-chain molecules (HC$_3$N and CCS) trace chemically young gas, while N-bearing molecules (N$_2$H$^+$ and NH$_3$) trace later stages of chemical evolution.
We investigate whether N$_2$H$^+$/HC$_3$N abundance ratio can be an indicator of the chemical evolution in the Vela C.
Figure \ref{column} shows the column density of N$_2$H$^+$ against that of HC$_3$N. 
We found a positive correlation between these column densities. We also found a systematic difference in the column densities between the star-forming and starless peaks.   
Figure \ref{fig1} shows the column density ratio of N$_2$H$^+$/HC$_3$N against the linewidth of HC$_3$N.
We find, on average, that the column density ratio of N$_2$H$^+$/HC$_3$N seems to be low toward starless peaks while it seems to be high toward star-forming peaks. The average value of the N$_2$H$^+$/HC$_3$N abundance ratio, excluding the data with upper limits, is 1.6$^{+0.5}_{-0.4}$ for star-forming (median value is 1.5) and 1.2$^{+0.4}_{-0.3}$ for starless peaks (median value is 0.9).  This is similar to the tendency found in the Orion A GMC and nearby cold dark clouds. \citet{oha14} suggested that column density ratio of NH$_3$/HC$_3$N decreases with increasing linewidth of HC$_3$N in the Orion A cloud. However, we found no correlation between  column density ratio of N$_2$H$^+$/HC$_3$N and linewidths. The criterion between star-forming and starless in the Vela C is found around N$_2$H$^+$/HC$_3$N $\sim$ 1.4. That is, N$_2$H$^+$/HC$_3$N may be $\lesssim$ 1.4 in starless peaks, and $\gtrsim$ 1.4 in star-forming peaks.  Differences in the filling factors of these two molecules may affect the column density ratio.  If the filling factor is equal to 0.5, the column density is $1.3-1.4$ times higher than that in the case of unity. Without having precise measurements of the filling factor, we assume the same filling factor, that is,  unity for both of them.  Even if the filling factor of HC$_3$N is systematically larger or smaller than that of N$_2$H$^+$, we will obtain the same tendency in comparison between starless and star-forming peaks, although
the absolute value of the ratio may change.
It should be noting that the uncertainties of the ratios are still large and we have to confirm these trends with high sensitivity observations in future.
\citet{tat14a} observed CCS $J=7-6$ and N$_2$H$^+$ $J=1-0$ emission toward dense cores in the Orion A cloud, and showed a clearer boundary between star-forming and starless at $\sim$ $2-3$ (see their figure 20). However, in the present study, the boundary between star-forming and starless is less evident. 
Our large beam observations may have failed to resolve the core. If so, the column density of HC$_3$N may be underestimated with our observations, and the column density ratio of N$_2$H$^+$/HC$_3$N will become smaller than our estimate in such cold dense regions. Furthermore, identifications of protostars may not be complete in the Vela C. \citet{gia12} identified protostars if 70 $\mu$m flux obtained by the $Herschel$ $Space$ $Observatory$ is more than 3 $\sigma$ from the best modified black body fit.  Embedded protostars may not be identified by these criteria and we may have misidentified star-forming cores as starless cores.
Finally, it may be also possible that the HC$_3$N $J=10-9$ transition is not a appropriate chemical evolution tracer as the CCS $J=7-6$ transition.  This is because the upper state energy for the HC$_3$N $J=10-9$ transition is $\sim$ 24 K, which is higher than that for the CCS $J=7-6$ transition of $\sim$ 15 K. The HC$_3$N $J=10-9$ transition may be emitted in warm gas in star-forming regions rather than in young cold gas.  It is also known that HC$_3$N emissions show the wing emission toward Orion KL 
\citep{ung97}  and  strong emission in the circumnuclear disk of NGC 1068 \citep{tak14}.  The HC$_3$N molecule will not trace always chemically young gas.

\begin{figure}[htbp]
  \begin{center}
  \includegraphics[width=12cm]{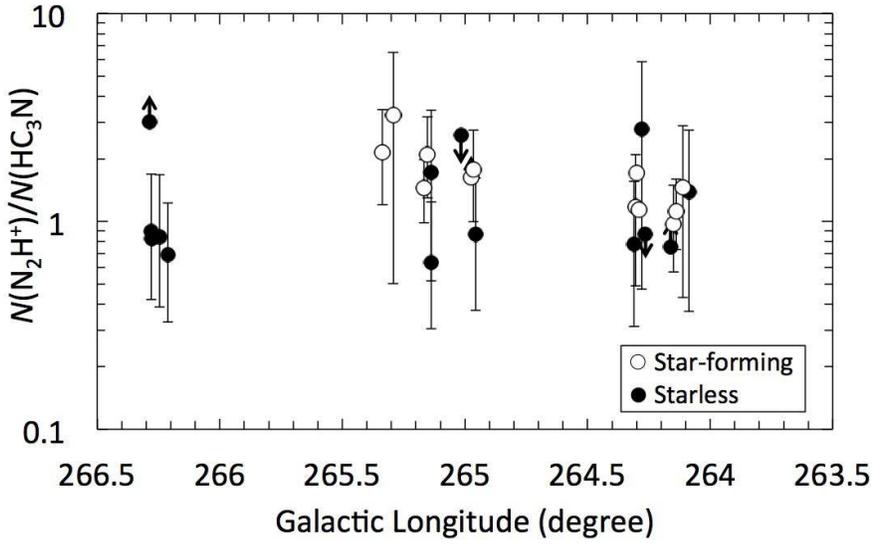}
  \end{center}
  \caption{The column density ratio of N$_2$H$^+$/HC$_3$N against the Galactic longitude.}
  \label{fig3}
\end{figure}

Finally, figure \ref{fig3} plots the column density ratio of N$_2$H$^+$/HC$_3$N against Galactic longitude. The lower longitude corresponds to the northern part of the Vela C and the higher longitude corresponds to the southern part.   The boundary between star-forming and starless cores may remain unchanged along the different locations of the cores.
We find no evidence for global chemical variation in Vela C, in contrast as has been observed in  Orion A GMC by \citet{tat10}.

\section{Summary}
We have observed the HC$_3$N $(J=10-9)$ and N$_2$H$^+$ $(J=1-0)$ lines toward the Vela C Molecular cloud. We found, on average, that  N$_2$H$^+$/HC$_3$N abundance ratio tends to be low in starless peaks while it tends to be high in star-forming peaks. This tendency is consistent with those found in the Orion A Cloud and dark clouds. The criterion between star-forming and starless found in the Vela C may be at $N$(N$_2$H$^+$)/$N$(HC$_3$N) $\sim$1.5. However, this separation is not clear compared with that of $N$(N$_2$H$^+$)/$N$(CCS) in the Orion A Cloud, suggesting either that our spatial resolution of 0.19 pc is insufficient to resolve spatial distribution differences between HC$_3$N and N$_2$H$^+$, the identification of protostars may be incomplete, or the HC$_3$N $J=10-9$ transition is not an appropriate chemical indicator in comparison with the CCS $J=7-6$ transition.\\
\\

 We thank the anonymous referee for helpful comments.
S. O. thanks Teresa Giannini for providing unpublished protostar data. Data analysis were carried out on common use data analysis computer system at the Astronomy Data Center, ADC, of the National Astronomical Observatory of Japan.

%


\end{document}